\documentclass[sigconf]{acmart}

\usepackage{subcaption}
\usepackage{soul}
\usepackage{multirow,rotating}
\AtBeginDocument{%
  \providecommand\BibTeX{{%
    \normalfont B\kern-0.5em{\scshape i\kern-0.25em b}\kern-0.8em\TeX}}}

\usepackage{acmart-taps}

\copyrightyear{2023}
\acmYear{2023}
\setcopyright{rightsretained}
\acmConference[CHIIR '23]{ACM SIGIR Conference on Human Information
Interaction and Retrieval}{March 19--23, 2023}{Austin, TX, USA}
\acmBooktitle{ACM SIGIR Conference on Human Information Interaction
and Retrieval (CHIIR '23), March 19--23, 2023, Austin, TX, USA}
\acmDOI{10.1145/3576840.3578312}
\acmISBN{979-8-4007-0035-4/23/03}

\begin{document}

\title{Why People Skip Music? On Predicting Music Skips using Deep Reinforcement Learning}

\author{Francesco Meggetto}
\affiliation{%
   \institution{NeuraSearch Laboratory, University of Strathclyde}
   \city{Glasgow}
   \country{UK}}
\email{francesco.meggetto@strath.ac.uk}

\author{Crawford Revie}
\affiliation{%
   \institution{University of Strathclyde}
   \city{Glasgow}
   \country{UK}}
\email{crawford.revie@strath.ac.uk}

\author{John Levine}
\affiliation{%
   \institution{University of Strathclyde}
   \city{Glasgow}
   \country{UK}}
\email{john.levine@strath.ac.uk}

\author{Yashar Moshfeghi}
\affiliation{%
   \institution{NeuraSearch Laboratory, University of Strathclyde}
   \city{Glasgow}
   \country{UK}}
\email{yashar.moshfeghi@strath.ac.uk}

\begin{abstract}
Music recommender systems are an integral part of our daily life. Recent research has seen a significant effort around black-box recommender based approaches such as Deep Reinforcement Learning (DRL). These advances have led, together with the increasing concerns around users’ data collection and privacy, to a strong interest in building responsible recommender systems. A key element of a successful music recommender system is modelling how users interact with streamed content. By first understanding these interactions, insights can be drawn to enable the construction of more transparent and responsible systems. An example of these interactions is skipping behaviour, a signal that can measure users’ satisfaction, dissatisfaction, or lack of interest. In this paper, we study the utility of users’ historical data for the task of sequentially predicting users’ skipping behaviour. To this end, we adapt DRL for this classification task, followed by a post-hoc explainability (SHAP) and ablation analysis of the input state representation. Experimental results from a real-world music streaming dataset (Spotify) demonstrate the effectiveness of our approach in this task by outperforming state-of-the-art models. A comprehensive analysis of our approach and of users’ historical data reveals a temporal data leakage problem in the dataset. Our findings indicate that, overall, users' behaviour features are the most discriminative in how our proposed DRL model predicts music skips. Content and contextual features have a lesser effect. This suggests that a limited amount of user data should be collected and leveraged to predict skipping behaviour.
\end{abstract}



\sloppy
\maketitle

\section{Introduction}
In recent years, online music streaming services (e.g., Spotify) have seen substantial growth. With the rise of digital music distribution, the related success of such streaming services, and the ubiquitous availability of music, a new listening paradigm has emerged. Users can access any song, at any time, and within a few clicks. As a result, there has been a significant change in users' behaviour and interaction with these systems \cite{fields2011contextualize}. Music recommender systems (MRS) aspire to tackle the problem of providing the users the support they need to access these large collections of music items and find songs that match their interests and needs. Recent research has seen a significant effort towards black-box based approaches such as Deep Reinforcement Learning (DRL) \cite{afsar2021reinforcement, sutton2018reinforcement}. This is motivated by the possible radical changes of behaviour from one song to another, or even within the same song, but at different points in time. Users' behaviour is influenced by external (trends) and internal (individual changes of personal interests) factors. The users' shifting interests and behaviour make it hard to learn a generalisable model to tailor the user's specific needs at any given time; it is a case where DRL is required due to continuous learning and adaption \cite{beutel2018latent, jannach2022session, wang2021survey}. These advances, however, have inevitably led to rising concerns about how users' data is collected, stored, and used. This is leading to a strong research interest in building responsible systems and data collection procedures \cite{milano2020recommender}. Constraints should be put in place when considering what data is collected and then presented to a model to measure user behaviours. This is due to the potential hazard of introducing errors and biases. Therefore, minimising and selecting high-quality data features is of important consideration. 

With explicit rating data relatively scarce and rare in today's systems, modelling implicit feedback is becoming of acquired importance. For example, in a \textit{lean-back} formulation, the case of automatic playlists or radio streaming, the user interaction is minimised. Users are presented with a single song at a time. The MRS needs to rely almost entirely on implicit feedback signals such as the skipping or scrubbing (i.e., seeking forward and backward by moving the cursor \cite{lamereScrub}) to predict satisfaction and engagement \cite{jannach2022session, wang2021survey}. By understanding these interactions, insights can be drawn for the construction of more transparent and responsible systems. The skipping is a signal that can measure users' satisfaction, dissatisfaction or lack of interest, and engagement with the platform \cite{meggetto2021skipping}. In a \textit{lean-back} formulation, the MRSs are often designed to be more conservative, prioritising \textit{exploitation} over \textit{exploration} to minimise negative feedback (in this context, skips) \cite{schedl2022music}. Thus, one of their goals may be determined as recommending songs that yield the highest listening activity (i.e. no skip). However, understanding the users' skipping behaviour is still an under-explored domain \cite{lamereSkip, brost2019music, meggetto2021skipping}. It is a challenging problem due to its noisy nature: a skip may suggest a negative interaction, but a user may skip a song that they like because they recently heard it elsewhere. In this work, we aim to understand why people skip by comprehensively analysing the utility of users' historical data. In particular, we analyse the impact and effect of the users' behaviour (e.g., the user action that leads to the current playback to start), listening content (i.e., the listened song), and contextual (e.g., the hour of the day) features in the classification task of predicting the users' music skipping behaviour. We propose a novel approach that leverages and adapts DRL for this classification task. This is to most closely reflect how a DRL-based MRS could learn to detect music skips.

Prior works in analysing the skipping behaviour revealed an universal behaviour in skipping across songs, with geography, audio fluctuations or musical events, and contextual listening information affecting how people skip music \cite{montecchio2020skipping, donier2020universality, ng2020investigating, meggetto2021skipping}. Recently, the effectiveness of deep learning models has also been explored for the task of predicting the users' sequential skipping behaviour in song listening sessions \cite{zhu2019session, hansen2019modelling, chang2019sequential, jeunen2019predicting, adapa2019sequential, tremlett2019preliminary, beres2019sequential}. While they made a significant contribution towards this direction, their process is usually seen as an independent and static procedure. They may not account for the dynamic nature of the users' behaviour, and do not intuitively optimise for the long-term potential of user satisfaction and engagement \cite{zheng2018drn, zhao2018deep, shani2005mdp, liu2018deep, jannach2022session, wang2021survey}. Overall, this motivates the investigation of the DRL's applicability in predicting music skips and a comprehensive investigation on the relation of the skipping signal with users' behaviour, listening context, and content. This paper aims to investigate the following two important research questions: \textit{can DRL be applied to the users' music skipping behaviour prediction task, and if so, would it be more effective in the music skip prediction task than deep learning state-of-the-art models?} (\textbf{RQ1}); \textit{what historical information is considered discriminative and serves as a high-quality indicator for the model to predict why people skip music?} (\textbf{RQ2}). To investigate our RQs, we have conducted an extensive study on a real-world music streaming dataset (Spotify). Our comprehensive analysis demonstrates the effectiveness of our approach and a temporal data leakage problem in the historical data. Overall, our findings indicate that the most discriminative features for our proposed DRL model to predict music skips are some users' behaviour features, with content and contextual features reporting a lesser effect. This suggests that a limited amount of user data can be leveraged to predict this behaviour, thereby offering implications in the building of novel user-centred MRSs and responsible data collection procedures. This is a necessary step in creating a holistic representation of the listeners' preferences, interests, and needs. The main contributions of this paper are:
\begin{itemize}
    \item We demonstrate the applicability and effectiveness of DRL in predicting users' skipping behaviour from listening sessions. A framework is devised to extend the DRL's applicability to perform this classification and offline learning. This is the first time that DRL has been explored in this task. The effectiveness of our approach is empirically shown on a real-world music streaming dataset (Spotify). Our proposed approach outperforms state-of-the-art models in terms of Mean Average and First Prediction Accuracy metrics.
    \item We perform a comprehensive post-hoc (SHAP) and ablation analysis of our approach to study the utility of users' historical data in detecting music skips. We reveal a temporal data leakage problem in the historical data. Further, our results indicate that overall users' behaviour features are the most prominent and discriminative in how the proposed DRL model predicts music skips. The listening content and context features are reported to have a lesser effect.
\end{itemize}

\section{Related Work}
\label{sec:relatedwork}

A successful MRS needs to meet the users’ various requirements at any given time \cite{song2012survey, hansen2020contextual, wen2019leveraging}. Thus, user modelling is a key element. A line of research has tried to untangle the relationship between personality and the users' musical preferences \cite{rentfrow2003re, rentfrow2006message, langmeyer2012music}. Volokhin and Agichtein \cite{volokhin2018understanding} introduced the concept of music listening intents and showed that intent is distinct from context (user's activity). A different, and arguably complementary, research direction is trying to understand and model how users interact with the underlying platform. This is a long-standing and under-researched problem of online streaming services \cite{brost2019music}. An example of these interactions is the skips between songs. Its modelling and understanding during music listening sessions plays a crucial role in understanding users' behaviour \cite{meggetto2021skipping}. The skips are often the only information available to the underlying MRS, and therefore they are used as a proxy to infer music preference \cite{schedl2022music}.

The skipping signal has already been used in prior works, as a measure in heuristic-based playlist generation systems \cite{bosteels2009evaluating, pampalk2005dynamic}, user satisfaction \cite{hansen2020contextual, wen2019leveraging}, relevance \cite{hansen2021shifting}, or as a counterfactual estimator \cite{mcinerney2020counterfactual}. Furthermore, given its universality and presence in other domains, recent research has also investigated its effect in ads on social media platforms \cite{banerjee2021skipping, belanche2017user, belanche2020brand}. Despite being abundant in quantity, it is a noisy implicit signal \cite{wang2021denoising, schedl2022music}. A skipped track does not necessarily imply a negative preference. Multiple hypotheses can be formulated on why users skip songs, with recent research suggesting that people manifest an universal behaviour in skipping across songs, dictated by time, geography, and reaction to audio fluctuations or musical events \cite{montecchio2020skipping, donier2020universality, ng2020investigating}. Moreover, it has been shown in \cite{taylor2021influence} that people who usually listen to songs in their entirety, show higher listening duration that those who do not. Most recently, Meggetto et al. \cite{meggetto2021skipping} proposed a clustering-based approach that clearly identifies four user types with regards to their session-based skipping activity. These types, namely \textit{listener}, \textit{listen-then-skip}, \textit{skip-then-listen}, and \textit{skipper}, are influenced by the length of the listening session, time of the day, and playlist type. The main limitation of these prior works is that they explore the relation between listening context and content with the skipping behaviour. They do not explore how the user interactions with the platform influence the detection of skips. This is a limitation that this works addresses.

In 2019, Spotify identified music skip prediction as an important challenge and organised the \textit{Sequential Skip Prediction Challenge} \footnote{\url{https://www.aicrowd.com/challenges/spotify-sequential-skip-prediction-challenge}} to explore approaches that could alleviate this problem. The challenge focused on predicting whether individual tracks encountered in a listening session will be skipped or not. To respond to this challenge, several deep-neural networks \cite{zhu2019session, hansen2019modelling, chang2019sequential, jeunen2019predicting, adapa2019sequential, tremlett2019preliminary, beres2019sequential} and supervised learning \cite{ferraro2019skip} models were proposed. Afchar and Hennequin \cite{afchar2020making} proposed using interpretable deep neural networks for skip interpretation via feature attribution. Whilst neural networks, and in particular Recurrent Neural Networks (RNNs), have been shown to effectively model sequential data, they consider the procedure as a static process. They do not intuitively provide a mechanism for the long-term optimisation of user satisfaction and engagement, continuous learning, and the modelling of the dynamic nature of the user's behaviour \cite{shani2005mdp, zheng2018drn, zhao2018deep, jannach2022session, wang2021survey}. Therefore, it is a case where DRL is required, an investigation and application of which has never been explored before. A research gap this work aims to address.

The \textit{Sequential Skip Prediction Challenge} is a binary classification task. Despite receiving limited attention to date, DRL has been shown to be suitable and effective in classification tasks. It can assist classifiers in learning advantageous features \cite{dulac2011datum, janisch2019classification} and select high-quality instances from noisy data \cite{feng2018reinforcement}. Wiering et al. \cite{wiering2011reinforcement} demonstrate that RL is indeed suitable for classification. Their model slightly outperforms existing classifiers, but training time and extra computational requirements are major drawbacks. With the recent advances in the field, a body of research is showing the superiority of DRL-based approaches for classification tasks \cite{janisch2019classification, janisch2020classification, martinez2018deep, feng2018reinforcement, lin2020deep}. In particular, the authors in \cite{janisch2019classification, janisch2020classification} show that a Vanilla Deep Q-Network (DQN) \cite{mnih2015human} approach is superior and more robust to state-of-the-art algorithms.

In this work, we explore, for the first time, the applicability of DRL in the task of sequentially predicting users' music skipping behaviour. This is motivated by the limitations of existing approaches and the advantages of DRL. By comprehensively analysing users' historical data, we study its utility and effect in our approach for this task. This work is the first step in understanding why people skip music.

\section{Approach}
\label{sec:approach}

In this section, we present our framework to facilitate the application of DRL to the problem of sequentially predicting users' skipping behaviour from listening sessions. To do so, we model this problem as a Markov Decision Process (MDP) and a mechanism is introduced in the RL problem formulation to correctly exploit logged interactions and thus perform offline learning. The details of this framework are as follows:

\noindent {\bf State}: it is the record-level representation of a listening session at a discrete time step (i.e., position in the session). The state, i.e. a record in a listening session, includes various user's contextual information about the stream, their interaction history with the platform, and information about the track that the user listened to. An episode is the entire listening session, with sessions containing from 10 up to at most 20 records.

\noindent {\bf Actions}: it is a discrete action space which is a binary indicator of whether the current track is going to be skipped or not by the corresponding user. Effectively, the problem formulation can also be thought of as a binary classification problem $A = \{0, 1\}$, where $0$ represents a no skip operation and $1$ represents a skip.

\noindent {\bf Reward}: a positive reward of 1 is given for a correctly predicted skip classification, 0 reward (i.e., no penalty) otherwise.

Motivated by the discrete action space and off-policy requirements of the music skip prediction task, we leverage DQN\footnote{Due to space limitations, we refer the readers to \cite{mnih2015human} for the necessary background and overview of the algorithm.}. These requirements preclude the use of algorithms such as Deep Deterministic Policy Gradient (continuous action space) and Proximal Policy Optimization (on-policy learning). Whilst the problem is formulated as an MDP, it is partially observable (POMDP) by definition. This is because only partial information about the listening context and of the user is available \cite{dulac2021challenges}. Hence, in our problem formulation, we consider MDP and POMDP to be equivalent. This means that we do not perform any further processing of the state representation (e.g., masking of some features).

This classification formulation can be seen as a guessing game, where a positive reward is given for a correct guess, and no penalty is given for an incorrect one. Long-term optimisation via discount factor $\gamma$ can be thought of as a way to correctly guess as many records in an episode as possible. Since there is a sequential correlation among records within an episode (i.e., a music listening session), a high $\gamma$ value should be used. This corresponds to optimisation on the total number of correct guesses in an episode (long-term) rather than optimisation on the immediate ones (short-term). By taking into account previous points in time and the past interactions with the environment, the DRL agent makes fully informed decisions.

\subsection{Offline Mechanism}
The DQN's standard training procedure is entirely online. Online learning is an iterative process where the agent collects new experiences by interacting with the environment, typically with its latest learned policy. That experience is then used to improve the agent's policy. However, exploiting logged data may be helpful and informative for the agent as a form of (pre)training. In offline learning (Batch RL \cite{lange2012batch}), the agent's task is instead defined as learning from a static dataset. Policies are learnt from logged data, and no interactions with the underlying environment are required. Whilst our prior formulation would work in an online learning setting, it presents a major problem when performing offline learning. A misclassification would cause a transition to a new state, which is, however, not part of the original trajectory and thus not represented in the dataset as well. The agent will generate and associate a (discounted) cumulative reward to a wrongly generated trajectory that is substantially different from the original. Thus, a pure offline algorithm has to exclusively rely on the transitions that are stored in the dataset provided in the input. From our initial formulation, we need to account for those out-of-distribution actions.

Within the definition of the reward function itself, the out-of-distribution, untruthful action is marked as invalid and, if sampled by the agent throughout learning, it causes the current episode to be terminated. In other words, an incorrect guess (0 reward) leads to a terminal state. This simple constraint forces a minimisation of estimation errors and therefore it avoids the creation of potential estimation mismatches. As such, the untruthful action that causes the current episode to terminate avoids the future propagation of incorrect bootstrapped return estimations in the Temporal Difference target. This is to minimise the distributional shift issues due to differences between the agent's policy and the behaviour policy. More specifically, it explicitly ensures that regardless of the next sampled action, the current policy $\pi (a' |s')$ is as close as possible to the behaviour distribution $\pi_{\beta} (a' |s')$. The Q-function is queried as little as possible on out-of-distribution and unseen actions since this will eventually increase errors in the estimations.

This error, i.e. "extrapolation error" \cite{fujimoto2019off}, is introduced when an unrealistic and erroneous estimation is given to state-action pairs. This is caused when action $a'$ from estimate $Q(s,a)$ is selected, and the consequent state-action pair $(s', a')$ is inconsistent with the dataset due to the pair being unavailable. It provides a source of noise that can induce a persistent overestimation bias and that cannot be corrected, in an off-policy setting, due to the inability to collect new data \cite{fujimoto2019off,fujimoto2018addressing}. Directly utilising DQN in an offline setting may result in poorer performance and a resemblance to overfitting \cite{levine2020offline}. Our proposed mechanism minimises these errors. It is important to note that the "correct" action is not forcefully fed to the agent as in Behaviour Cloning based approaches. We let the agent deterministically decide as if it were a live interaction with the environment, thus keeping the general workflow of the original algorithm intact. This provides a single interface to easily transition from offline to online learning and vice versa.

Finally, it is important to note that the aim of this work is to enhance our understanding of why people skip music and identify the high-quality features for its detection. To this end, we analyse the applicability of DRL in predicting this behaviour. We leave further tailoring of the approach to the music skip prediction task and an evaluation with recently proposed offline model-free algorithms \cite{fujimoto2019benchmarking, agarwal2020optimistic, dabney2018distributional, kumar2019stabilizing} for future work. Nevertheless, our proposed approach requires no architectural or algorithmic modifications. It offers the potential for a swift transitioning from online to offline learning and vice versa. It can be also be considered as a swift pre-training of an agent that can later be deployed online for continual learning.

\section{Experimental Settings}
\label{sec:experimentalSettings}

\subsection{Dataset}
\label{sec:dataset}
We conduct our experiments on the real-world Music Streaming Sessions Dataset (MSSD) provided by Spotify \cite{brost2019music}. The publicly available training set consists of approximately 150 million logged streaming sessions, collected over 66 days from July 15th and September 18th 2018. Each day comprises ten logs, where each log includes streaming listening sessions uniformly sampled at random throughout the entire day. Sessions contain from 10 up to at most 20 records and are defined as sequences of songs/tracks that a user has listened to (one record per song). Each record includes various user's contextual information about the stream (e.g., the playlist type) and interaction history with the platform (e.g., scrubbing, which is the number of seek forward/back within the track). Although the track titles are not available, descriptive audio features and metadata are provided for them (e.g., acousticness, valence, and year of release). It is important to note that there is no user identification, nor access to demographic or geographical information. Hence, by not knowing whether two sessions have been played by the same user or by two different users, this study revolves around the modelling and understanding of the users' skipping behaviour.

\subsubsection{Temporal Correlation}
There is no temporal correlation among listening sessions, i.e. the sessions are not presented in historical order, which is reflected in the chance of consecutive sessions having a considerably different hour of the day (e.g., morning and evening). Also, there is no order to the ten logs within a given day (i.e., the 1st log of the first day does not necessarily occur before the 2nd of the same day). This does not preclude the potential applicability of DRL for the skip prediction task since the hour of the day in which a song was played is provided. Thus, it allows for the modelling of skipping behaviour dependent on the hour of the day.

\begin{table*}
    \caption{Summary of datasets used for experiments after pre-processing. log(s) \# indicate which log(s) are selected out of the available ten. skip (\%) refers to the ratio between True and False values.}
    \label{tab:datasetStats}
    \begin{tabular}{c c c c c c}
        \toprule
        Dataset & Date & log(s) \# & \# of records & \# of sessions & skip (\%) \\
        \midrule
        Training Set     & 15/07/2018 & [0, 3] & 11,927,861 & 711,838 & 51.20\% \\
        Test Set (T1) & 15/07/2018 & 4 & 2,991,438 & 178,419 & 51.21\% \\
        Test Set (T2) & 19/07/2018 & 8 & 3,395,883 & 204,145 & 50.53\% \\
        Test Set (T3) & 27/07/2018 & 0 & 3,447,209 & 207,060 & 50.76\% \\
        Test Set (T4) & 10/08/2018 & 6 & 3,407,685 & 205,267 & 50.42\% \\
        Test Set (T5) & 09/09/2018 & 1 & 2,588,711 & 155,617 & 51.48\% \\
        \bottomrule
\end{tabular}
\end{table*}

\subsubsection{Creation of Training and Test Sets}
\label{sec:creationTrainTestSet}
In this work, we only leverage the training set since, in the test set, most of the metadata and the skipping attributes used as ground truth in our evaluation are not provided. By selecting logs from the original training set, statistics for our training and test datasets are presented in Table~\ref{tab:datasetStats}. As it can be seen from the statistics, the ratio of skip values for all sets is balanced between True and False values. This balanced distribution is an intrinsic property of the dataset and of any of the available logs. Due to the large amount of data, and therefore computational and execution time requirements, the first four logs of the first available day are used for training. Testing is performed on various logs in order to test the models' generalisability for different days. Except for T1, which is the 5th and next immediate consecutive log after the training set collection, all the other logs are of a random index, day and/or month. This random selection approach is justified by the fact that there is no temporal correlation among logs of the same day. This is to show the generalisation capabilities of our proposed approach and to allow for the comprehensive analysis of the importance of the users' historical data.

\subsubsection{Data Preprocessing}
\label{sec:dataPreprocessing}
All available features, with a full description available in \cite{brost2019music}, are included in the state representation, except for the skip features, session and song identifiers. Categorical features, such as the playlist type and the user's actions that lead to the current track being played or ended, are one-hot encoded. All the audio features are standardised to have a distribution with a mean value of 0 and a standard deviation of 1. Overall, this results in a state representation consisting of 70 features. For ease of discussion, they are grouped as follows:

\noindent {\bf User Behaviour (UB)}:
\begin{itemize}
    \item \textbf{Reason End (RE)} is the cause of the current playback to end. This is a one-hot encoded feature that thus groups various encoded features such as \textit{Trackdone}, \textit{Backbtn}, \textit{Fwdbtn}, and \textit{Endplay}.   
    \item \textbf{Reason Start (RS)}. Similar to \textit{Reason End}, it is the type of actions that cause the current playback to start.
    \item \textbf{Pauses (PA)} is the length of the pause in between playbacks. It consists of \textit{No}, \textit{Short}, and \textit{Long Pause}.
    \item \textbf{Scrubbing (SC)} is the number of seeking forward or backward during playback. They correspond respectively to \textit{Num Seekfwd} and \textit{Num Seekback}.
    \item \textbf{Playlist Switch (PS)} indicates whether the user changed playlist for the current playback.
\end{itemize}

\noindent {\bf Context (CX)}:
\begin{itemize}
    \item \textbf{Session Length (SL)} is the length of the listening session.
    \item \textbf{Session Position (SP)} is the position of the track within the session.
    \item \textbf{Hour of Day (HD)} is the hour of the day in which the playback occurred ($[0..23]$).
    \item \textbf{Playlist Type (PT)} is the type of the playlist that the playback occurred within. Examples are \textit{User Collection}, \textit{Personalized Playlist}, and \textit{Radio}.
    \item \textbf{Premium (PR)} indicates whether the user was on premium or not.
    \item \textbf{Shuffle (SH)} indicates whether the track was played with shuffle mode activated.
\end{itemize}

\noindent {\bf Content (CN)}. This third and final category groups all the \textbf{Track (TR)} metadata and features, as they constitute the only content-based information in the MSSD. It includes 28 features such as \textit{Beat Strength}, \textit{Key}, \textit{Duration}, and the eight \textit{Acoustic Vectors} ([0..7]).

\subsection{Evaluation Metrics}
\label{sec:evalMetrics}
To perform an evaluation of our proposed approach, we adopt the evaluation metrics from the \textit{Spotify Sequential Skip Prediction Challenge}. This is also to provide a fair comparison with the selected baselines, since they were proposed on this challenge and for the following task: \textit{given a listening session, predict whether the individual tracks encountered in the second half of the session will be skipped by a particular user.} Therefore, every second half of a session in the selected test set is used for prediction. If a session has an odd number of records, the mid-value is rounded up. This is motivated by the fact that an accurate representation of the user's immediately preceding interactions can inform future recommendations generated by the music streaming service. Hence, it is important to infer whether the current track is going to be skipped as well as subsequent tracks in the session. First Prediction Accuracy and Mean Average Accuracy are adopted as metrics.

{\bf \noindent First Prediction Accuracy (FPA)} is the accuracy at predicting the first interaction for the second half of each session.

{\bf \noindent Mean Average Accuracy (MAA)} is defined as: 
\begin{equation}
    MAA = \frac{\sum\limits_{i=1}^T A(i) L(i)}{T}
\end{equation}
where $T$ is the number of tracks to be predicted within the given session, $A(i)$ is the accuracy up to position $i$ of the sequence, and $L(i)$ indicates whether the $i^{th}$ prediction is correct or not. Intuitively, in these evaluation metrics higher importance is given to early predictions. In our setting, however, we do not exploit this specification in the problem formulation. Instead, the agent is instructed to optimise the total number of correct predictions in the session. This is to keep the system's specifications simple and easily adaptable to different metrics and/or tasks. In the dataset schema, prediction is based on the \textit{skip\_2} feature. It indicates a threshold on whether the user played the track only briefly (no precise threshold is provided) before skipping to the next song in their session.

\subsection{Models}
\subsubsection{Baselines}
\label{sec:baselines}
To identify state-of-the-art baselines on the music skip prediction task, we performed an extensive search on prior works that utilise the MSSD dataset. We identified the following 4 of the top-5 ranked submissions to the
\textit{Spotify Sequential Skip Prediction Challenge} and presented at the WSDM Cup 2019 Workshop:
\begin{itemize}
    \item \textbf{Multi-task RNN}: RNN-based approach that predicts multiple implicit feedbacks (multi-task) \cite{zhu2019session}.
    \item \textbf{Multi-RNN}: Multi-RNN with two distinct stacked RNNs where the second makes the skip predictions based on the first, which acts as an encoder \cite{hansen2019modelling}.
    \item \textbf{Temporal Meta-learning}: A sequence learning, meta-learning, approach consisting of dilated convolutional layers and highway-activations \cite{chang2019sequential}.
    \item \textbf{Weighted RNN}: RNN architecture with doubly stacked LSTM layers trained with a weighted loss function \cite{jeunen2019predicting}.
\end{itemize}

They respectively reported the 1st, 2nd, 3rd, and 5th best overall performance on the Spotify Challenge, with Multi-task RNN being the strongest and Weighted RNN being the weakest baselines. The exclusion of the 4th overall best model on the challenge in our evaluation is because no manuscript and code repository were found. For the selected baselines, we use the code accompanying the papers (GitHub links available in cited manuscripts). We then reproduced their results locally by running their provided public code locally, to the best of our abilities and with an optimised set of parameters. However, despite our best efforts, we reported consistently worse results than the ones in the Spotify Challenge public leaderboard and/or accompanying papers. The test set used in challenge is not fully released. No ground truth is available, thereby not allowing for a local evaluation. However, given our procedure for the creation of the train and test sets (Section \ref{sec:creationTrainTestSet}), i.e. the training is performed on the first available day and the evaluation is for different days/months, we make the strong assumption that the overall data distribution of our selected test sets and the one used in the public challenge are similar. For a fair comparison, we thus report the results from the public leaderboard since they are better than the ones from our local evaluation.

\subsubsection{DQN Architecture}
For this work, we explored nine state-of-the-art DQN architectures. By adhering to our proposed framework, they have been thoroughly investigated in the users' music skipping behaviour prediction task. They are the Vanilla \cite{mnih2015human}, Double \cite{van2016deep}, Dueling \cite{wang2016dueling}, and their respective n-step learning variants \cite{mnih2016asynchronous}. Partially observable architectures have also been explored, with observations stacking \cite{mnih2015human} and Gated Recurrent Units (GRU) and Long Short-Term Memory (LSTM) based architectures \cite{hausknecht2015deep}.

Due to space limitations, a comparison among all these architectural variants is not reported. We note, however, that Vanilla DQN achieves the best performance. This is given its comparable performance and the advantage of a significantly simpler architecture with lower complexity. Therefore, the reported results are only for the Vanilla DQN architecture (hereafter referred to as "DQN").

\subsection{Experimental Procedure}
We trained our DQN using the following set of parameters: experience replay memory is 10000, batch size and frequency of updates are set as 256, the learning rate is 0.001, and the discount factor is 0.9. The policy network consists of three fully-connected layers (of size 128) and a final action-value linear output layer of size 2. This final layer computes the Q-values for each action. Hyperparameters were selected by random and Tree-structured Parzen Estimator search, with the best set selected for evaluation on the test collections. The implementation of the DQN agent is provided by the Tensorforce \cite{tensorforce} library. For complete reproducibility of our work, the code for this work is available at \url{https://github.com/NeuraSearch/Spotify-XRL-Skipping-Prediction}

To explore the potential instabilities and divergences during training, the proposed DQN approach is run five times per test set. The reported results represent the mean across all test sets. Lastly, during the training phase, learning is constrained with out-of-distribution actions, and therefore, some state-value pairs in the dataset are not experienced by the agent due to early termination. During the testing phase, all episodic records are sequentially retrieved, and the agent acts deterministically on the complete episodes for its evaluation.

\subsubsection{Post-Hoc Analysis}
In order to carry out an analysis on the importance and validity of the users' historical data in predicting music skipping behaviour, we first leverage the Shapley Additive Explanations framework (SHAP) \cite{lundberg2017unified}. It is a game-theory based approach that explains the predictions of machine learning models. In particular, we adopt the Kernel Explainer, which is a model agnostic method to estimate the SHAP values. This is because there exists no DRL specific explainers. However, since the Kernel Explainer makes no assumptions about the model to explain the predictions of, it is a highly expensive computational approach. This means that it is slower than the other model type specific algorithms. By considering these extensive computational requirements, for each test set, we estimate the feature importance values for the first 50 episodes (i.e., listening sessions) and with 200 perturbation samples per record. Given the high similarity across all test sets, we only report the results for T1.

\subsubsection{Ablation Analysis}
To validate the SHAP results, we perform an ablation analysis on the input state representation. We study the effect that the category (e.g., \textit{UB}) and type (e.g. \textit{RS}) features have on the DQN's performance. To this end, we train and evaluate (following the same above-mentioned experimental procedure) the proposed approach on a state representation that does not include the selected features' type. This iterative approach, whereby only a single type is removed for each evaluation, is repeated until all types that comprise the input state representation are evaluated.

\subsubsection{Temporal Data Leakage}
\label{sec:dataLeakage}
A closer investigation of the MSSD dataset, and validated by the post-hoc and ablation analysis, reveals a temporal data leakage of some features. These features have been left unnoticed and they have inadvertently affected the Spotify Challenge and thus the baselines. These features correspond to the length of session (\textit{SL}) and the user actions that lead the current playback to end (\textit{RE}). This is because they provide to the model information from the future that should not be available in a live predictive system. Although we recognise and acknowledge this to be a problem, the reported results on the comparison with the selected baselines are without the removal of such features. This is to provide a fair comparison with the selected baselines, since they include these features in their input representation. These features are removed from the state when we investigate why people skip music, and it is referred to as the "corrected" state.

\section{Results}
\label{sec:results}

First, the validity of our approach to predict users' music skipping behaviour is demonstrated against the state-of-the-art deep learning based models. Our analysis of the music skipping prediction task and of the MSSD dataset reveals a temporal data leakage problem (Section \ref{sec:dataLeakage}). With a "correction" of the state representation by removal of such features, we report the comprehensive investigation on how the skipping behaviour can be detected by analysing the importance of UB, CX, and CN.

\subsection{Applicability of DRL to Music Skip Prediction (RQ1)}
\label{sec:resultsDRLApplicability}
\begin{table*}
    \caption{MAA and FPA results for our proposed DQN approach and baselines. The reported results are the averages across all test sets for DQN (with 95\% CI). For the baselines, we report the publicly available results from the Spotify Challenge\protect\footnotemark. This is to provide a fair comparison since they are better than those obtained from our local evaluation. No CIs are reported for the baselines due to their unavailability. The best performing model is highlighted in bold.}
    \label{tab:tableDQNValidity}
    \begin{tabular}{@{}cl|lc|lc@{}}
        \hline
        & & \multicolumn{2}{c|}{\textit{MAA}} & \multicolumn{2}{c}{\textit{FPA}} \\
        \hline
        & & Mean & 95\% CI & Mean & 95\% CI  \\
        \hline
        & DQN & \textbf{0.820} & [0.818 - 0.822] & \textbf{0.881} & [0.880 - 0.882] \\
        \hline
        \multirow{-2}{*}{\aptLtoX[graphic=no,type=html]{Public Leaderboard}{\rotatebox{270}{\parbox[t]{1mm}{Public\\Leaderboard}}}} & Multi-task RNN & 0.651 & \textemdash & 0.812 & \textemdash \\
        & Multi-RNN              & 0.641 & \textemdash & 0.807 & \textemdash \\
        & Temporal Meta-learning & 0.637 & \textemdash & 0.804 & \textemdash \\
        & Weighted RNN           & 0.613 & \textemdash & 0.794 & \textemdash \\
        \hline
\end{tabular}
\end{table*}

\footnotetext{Leaderboard results available on cited manuscripts and/or at \url{https://www.aicrowd.com/challenges/spotify-sequential-skip-prediction-challenge}}

On our local evaluation, Multi-RNN and Temporal Meta-learning, despite outperforming Weighted RNN in the challenge submissions, perform consistently worse on our selected test sets. Multi-Task RNN, the best performing baseline on the public challenge, achieves slightly inferior performance compared to Weighted RNN. Overall, we note that all the baselines perform consistently worse on our local evaluation than in the public challenge. We observe decreases in performance of 4.9, 16.2, 4.9, 0.8 (\%) and 2.4, 8.2, 2.2, 0.4 (\%) in MAA and FPA and for Multi-task RNN, Multi-RNN, Temporal Meta-learning, and Weighted RNN respectively. Therefore, in Table~\ref{tab:tableDQNValidity}, we report results in terms of MAA and FPA metrics for our proposed DQN approach with the baselines' public results from the Spotify Challenge. This is because they are better than those that we obtained from our local evaluation and to provide an as fair as possible comparison. Our proposed approach exhibits significant improvements over all baselines on both MAA and FPA metrics. Our proposed DQN registers an increase of performance for both MAA and FPA of $17\%$ and $7\%$ respectively with regards to Multi-task RNN, the best performing baseline from the public challenge.

Overall, our results demonstrate the validity and applicability of DRL to predict users' music skipping behaviour. A Vanilla DQN architecture can outperform the more complex deep learning based state-of-the-art models. Furthermore, the results and a thorough analysis, omitted from this paper due to space limitations, also indicate that convergence is achieved using a significantly lower number of episodes, at around $2 \times 10^5$ ($\sim 1/4$ of the episodes in the training set). This suggests sample efficiency and swift convergence of our proposed approach. Thus, it also addresses the well-known problem of DRL, which is its computationally intensive and slow learning. Our approach converges swiftly and, in contrast to the selected baselines, it does not require GPU access. The low variability in performance across multiple runs and during the learning process also indicates stable and effective learning.

\subsection{Identification of Temporal Data Leakage}
In the previous section, we compared our proposed DQN against the selected baselines in order to demonstrate the validity of our proposed DQN. By performing an as fair as possible comparison, empirical results indicate the superiority of our approach. However, this benchmarking introduced errors into the model. This is because, as described in Section \ref{sec:dataLeakage}, we recognise that there are data leaking features in MSSD. The \textit{SL} informs the model of how many songs a given user will listen to. This should not be made available because it is impossible to know how many songs a user will listen to in their current listening session. Further, the \textit{RE} features provide information about how the current stream ends. This information should also not be exposed to the model. However, to provide a fair comparison with the baselines, since they are included in their input representation, these features were not removed despite our acknowledgement.

\begin{figure*}
    \centering
    \includegraphics[width=\textwidth]{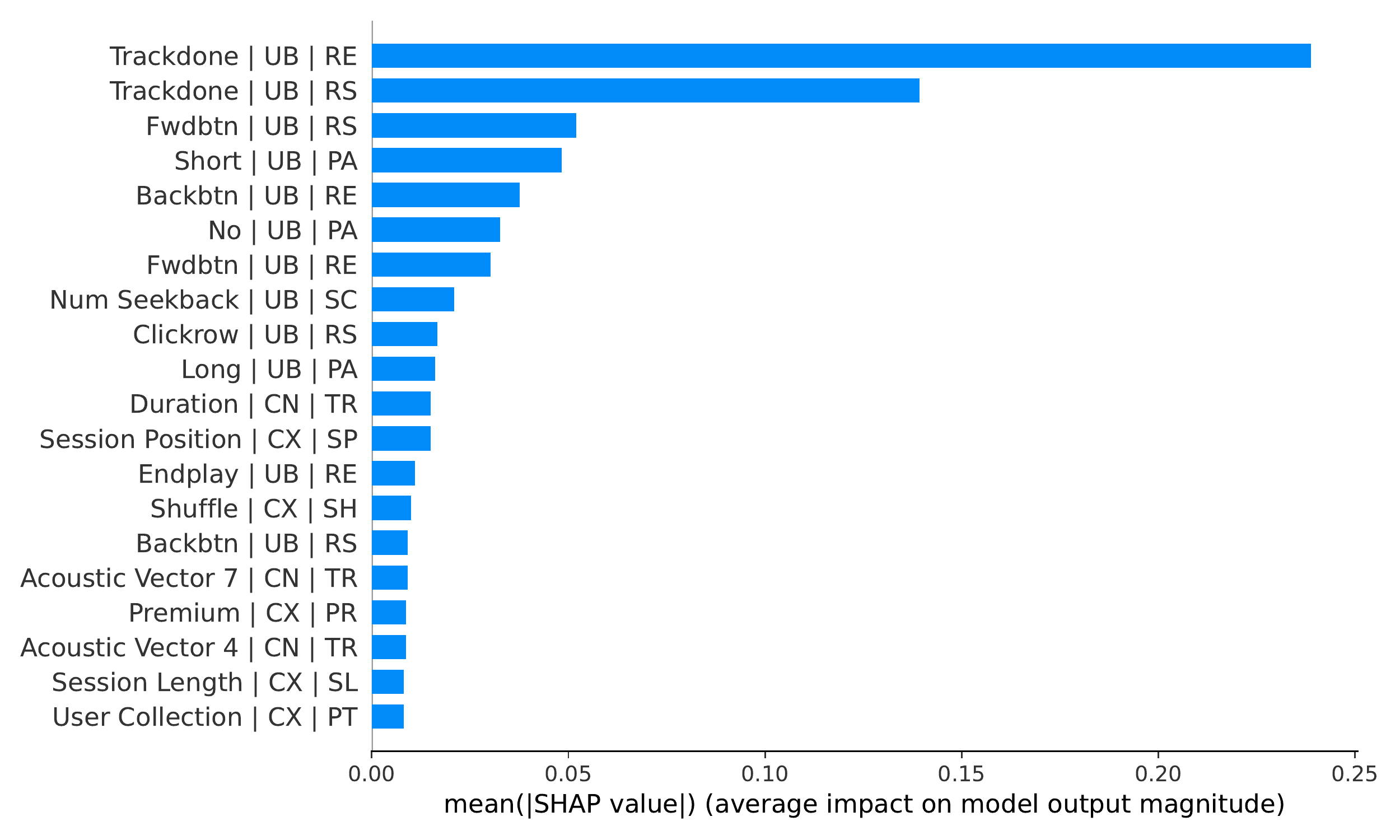}
    \caption{SHAP features importance analysis of the proposed DQN. The categorisation of the features and an explanation of the used acronyms is described in Section \ref{sec:dataPreprocessing}. Features are ranked in order of importance and they are reported as "[Name] | [Category] | [Type]".}
    \label{fig:figSHAPBefore}
\vspace*{-10pt}\end{figure*}

\begin{figure*}
    \centering
    \includegraphics[width=\textwidth]{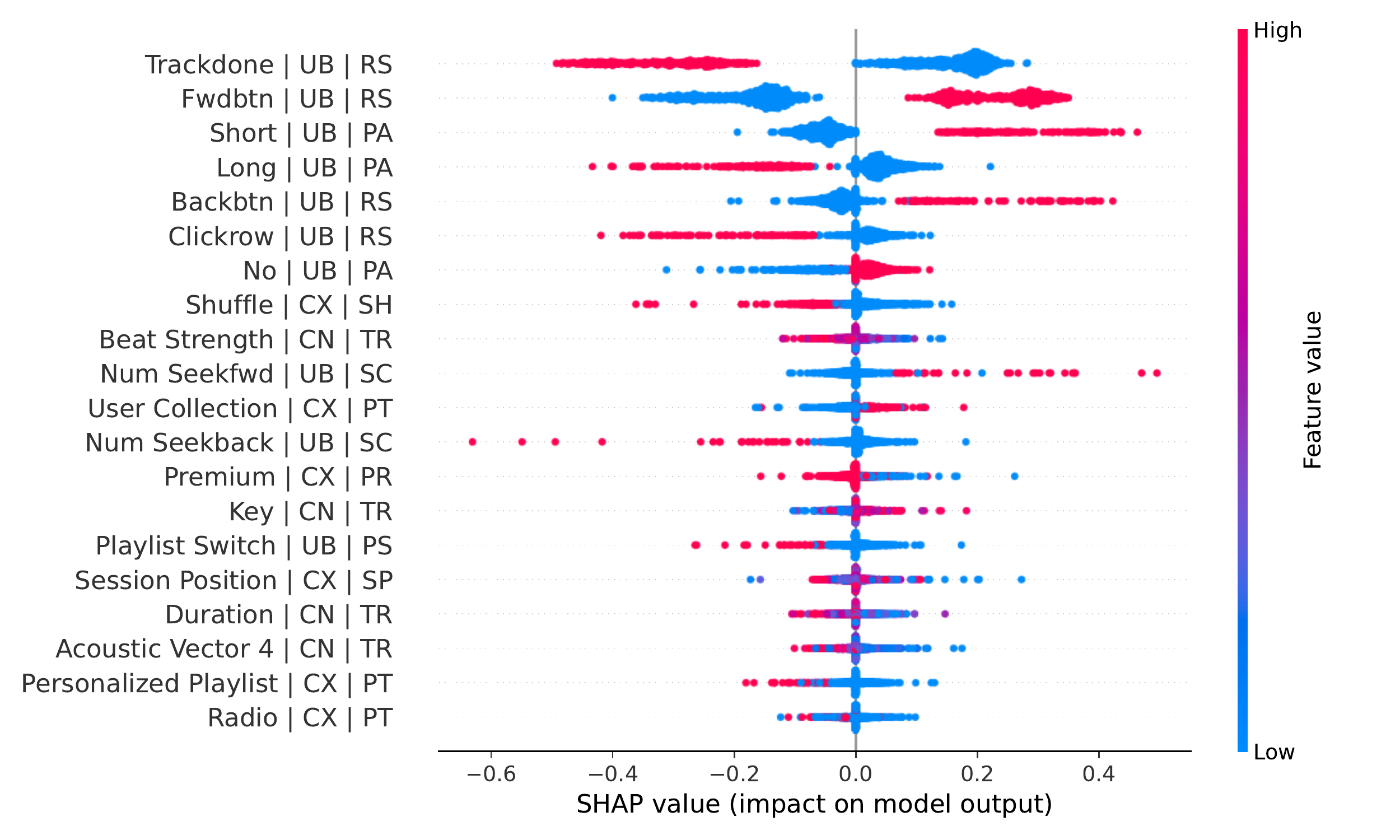}
    \caption{SHAP features importance analysis with positive (skip) and negative (no skip) impact values of the proposed DQN on a "corrected" state representation (i.e., after addressing temporal data leakage). The Feature Value axis refers to high or low observational values. For Boolean features (e.g., \textit{RS Trackdone}), high/red is a True value, and low/blue is False. The categorisation of the features and an explanation of the used acronyms is described in Section \ref{sec:dataPreprocessing}. Features are ranked in order of importance and they are reported as "[Name] | [Category] | [Type]".}
    \label{fig:figSHAPAfter}
\end{figure*}

The temporal data leakage problem is validated by Figure \ref{fig:figSHAPBefore}, which reports the analysis of the average impact on model output (SHAP) of all features in the input state representation. It can be noted how the most discriminative feature to detect music skips is \textit{RE Trackdone}, followed by \textit{RS Trackdone}, \textit{RS Fwdbtn}, and \textit{Short PA}. \textit{SL} is also found to have a relative impact (19th). It is clear that the proposed DQN considers these features to be of high quality and prominent importance for predicting the users' music skipping behaviour. However, they introduce a data leaking problem. By their removal from the input state representation, we observe a decrease in performance for our proposed DQN of $16\%$ and $11\%$ in MAA and FPA respectively. Further, we observe decreases in performance of 5.2, 26.2, 7.6, 0.6 (\%) and 3.5, 28.4, 6.0, 1.3 (\%) in MAA and FPA for Multi-task RNN, Multi-RNN, Temporal Meta-learning, and Weighted RNN respectively (differences calculated from the results obtained in our local evaluation after removal of the features with those reported in the public challenge). Overall, these results validate our initial intuition and demonstrate the data leakage problem. This finding provides a strong implication for a future outlook on creating attentive data collection procedures for transparent measurements of user behaviours. Offline benchmarks should be an as truthfully as possible reflection of real-world (online) tasks.

\subsection{The Role of User Behaviour, Context, and Content in Detecting Music Skips (RQ2)}
In this final section, we aim to address our main research question: why people skip music? To this end, we acknowledge and thus remove the leaking features from the state representation to enable for a correct modelling of the users' music skipping behaviour.

\subsubsection{\textbf{User Behaviour (UB)}}
Figure \ref{fig:figSHAPAfter} reports the SHAP features importance analysis of the proposed DQN on the "corrected" state representation. It can be observed that how the user interacted with the underlying platform to start the current playback (i.e., the \textit{RS} type) is considered being the most discriminative feature to detect music skips. \textit{Trackdone} and \textit{Fwdbtn} are the highest negatively and positively correlated features in predicting a skip. They correspond to the user starting the current playback having listened in full or having pressed the forward button (i.e., skip) on the previous playback. These findings validate the recent observations by Meggetto et al. \cite{meggetto2021skipping}. By considering their defined listener and skipper user types, we hypothesise that the user behaviour that can inform the membership of a user to one of these two types is a \textit{RS Trackdone} or \textit{Fwdbtn}. From our results, it is clear that how a person interacted with the previous song appears to greatly affect the DRL's ability to detect how they will interact next. Another UB that appears to have a prominent effect is the pause in between playbacks. A \textit{Short PA} and a \textit{No PA} are shown to highly and weakly suggest a music skip respectively. In the case of a \textit{Long PA}, our results strongly indicate that the user will not skip their current song. This finding validates our initial hypothesis. It may correspond to a person searching the catalogue for a song they would like to listen, and hence a long pause. Therefore, it is intuitive that it may not be skipped. However, the effect of a short pause in detecting music skips is of surprising effect. This may be justified by a user's exploratory state where they browse the catalogue and briefly listen to multiple songs until they find a match for their needs.

\subsubsection{\textbf{Context (CX)}}
We observe that users that listen in \textit{Shuffle} mode and/or with a \textit{Premium} account are associated with less skipping activity. Listening with a \textit{User Collection} PT is associated with a higher skipping rate. It is also shown that listening under a \textit{Personalised Playlist} or \textit{Radio} is subject to more listening and thus less skipping activity. This finding could suggest that they have a higher users' engagement. However, this is not possible to quantify, and further evaluation is required in order to understand this phenomenon. This could be explained by the noisy nature of the skipping activity and the possibility, as in the example of radio listening, of passive (background) consumption of the music. Although the \textit{PT} findings appear to partially validate prior work \cite{meggetto2021skipping}, in our ablation analysis we see that their removal from the state representation registers no significant effect on the DRL's ability to predict music skips.

\begin{table*}
    \caption{MAA and FPA results for our ablation analysis on the proposed DQN on the corrected state representation. The reported results are the average across all test sets and the 95\% CIs. (\textbf{*}) and (\textbf{**}) indicate that the selected type of features had a statistically significant effect in performance in the proposed DQN (on a "corrected state") on MAA or FPA. This is based on confidence levels ($p < .05$) and ($p < .001$) respectively.}
    \label{tab:ablationAnalysis}
    \begin{tabular}{@{}cl|lc|lc@{}}
        \hline
        & & \multicolumn{2}{c|}{\textit{MAA}} & \multicolumn{2}{c}{\textit{FPA}} \\
        \hline
        & & Mean & 95\% CI & Mean & 95\% CI  \\
        \hline
        & Corrected State & 0.664 & [0.662 - 0.666] & 0.773 & [0.772 - 0.774] \\
        \hline
        {\multirow{4}{*}{\aptLtoX[graphic=no,type=html]{UB}{\rotatebox{270}{UB}}}} & Reason Start (RS) & 0.389 (\textbf{**}) & [0.378 - 0.400] & 0.479 (\textbf{**}) & [0.464 - 0.494] \\
        & Pauses (PA) & 0.659 (\textbf{*}) & [0.657 - 0.661] & 0.769 (\textbf{*}) & [0.768 - 0.770] \\
        & Scrubbing (SC) & 0.659 & [0.655 - 0.663] & 0.770 (\textbf{*}) & [0.768 - 0.772] \\
        & Playlist Switch (PS) & 0.662 & [0.659 - 0.665] & 0.773 & [0.772 - 0.774] \\
        \hline
        \multirow{3}{*}{\aptLtoX[graphic=no,type=html]{CX}{\rotatebox{270}{{CX}}}} & Hour of Day (HD) & 0.663 & [0.661 - 0.665] & 0.773 & [0.772 - 0.774] \\
        & Playlist Type (PT) & 0.663 & [0.661 - 0.665] & 0.772 & [0.771 - 0.773] \\
        & Premium (PR) & 0.664 & [0.662 - 0.666] & 0.773 & [0.772 - 0.774] \\
        & Shuffle (SH) & 0.663 & [0.660 - 0.666] & 0.774 & [0.773 - 0.775] \\
        \hline
        \addlinespace[3pt]
        \multirow{-2}{*}{\aptLtoX[graphic=no,type=html]{CN}{\rotatebox{270}{{CN}}}} & Track (TR) & 0.664 & [0.661 - 0.667] & 0.773 & [0.772 - 0.774] \\
        \hline
    \end{tabular}
\end{table*}

\subsubsection{\textbf{Content (CN)}}
The only content-based features in the MSSD are related to the track being listened by the user (\textit{TR}). The correlation between skipping activity and the \textit{TR} features is less obvious since they appear to be less discriminative and promiment in detecting music skips. \textit{Beat Strength} and \textit{Key}, although mostly centred around a zero impact, suggest that a high beat strength is associated with more listening, and a high-pitched song (\textit{Key}) with higher chances of skipping. Further, longer songs (\textit{Duration}) are usually associated with higher listening activity, although they may also correspond to skips. However, in our ablation analysis, we observe the no effect in the DQN's performance by the removal of all \textit{TR} features. We find this to be of surprising effect, since it appears to contradict prior research suggesting that audio characteristics influence how people skip music \cite{donier2020universality, ng2020investigating}. 

\subsubsection{Ablation Analysis}
In order to validate our findings and to demonstrate the impact, whether statistically significant or not, that these features have on the DQN's performance, in Table \ref{tab:ablationAnalysis} we report the results for the ablation analysis. We performed paired t-tests on the prediction accuracy of the proposed DQN (on the "corrected" input state representation) with each of the selected type of features (e.g., \textit{RS}). We use (\textbf{*}) and (\textbf{**}) to denote the fact that the removal of the selected type of features had a statistically significant effect in performance in the proposed DQN on MAA and FPA. This is based on confidence levels ($p < .05$) and ($p < .001$) respectively. We note how the \textit{RS} features type, as previously shown in Figure \ref{fig:figSHAPAfter}, is the highest quality estimator to detect music skips. Its removal registers a decrease in performance of $28\%$ and $29\%$ in MAA and FPA respectively. The \textit{PA}s also register a significant impact. All the remaining features, including the CX and CN categories, do not appear to show a statistically significant effect on the DQN's performance. These results, therefore, suggest that a limited amount of users' data can be indeed leveraged to predict the users' music skipping behaviour, with only the \textit{RS} and \textit{PA} user behaviours showing a statistically significant effect.

\section{Discussion \& Conclusions}
\label{sec:conclusion}
In this work, we aim to understand why people skip music. To carry out such an analysis, we first proposed to leverage DRL to the task of sequentially predicting users' skipping behaviour in song listening sessions. By first understanding how a DRL model learns individual user behaviours, we can then help the process of explaining recommendations of a DRL-based MRS. To this end, we extended the DRL's applicability to this classification task. Results on a real-world music streaming dataset (Spotify) indicate the validity of our approach by outperforming state-of-the-art deep learning based models in terms of MAA and FPA metrics (\textbf{RQ1}). By empirically showing the effectiveness of our proposed approach, our main post-hoc and ablation analysis revolves around a comprehensive study of the utility and effect of users' historical data in how the proposed DRL detects music skips (addressing \textbf{RQ2}).

Our findings indicate that how users interact with the platform is the most discriminative indicator for an accurate detection of skips (i.e., \textit{RS} and \textit{PA}). Surprisingly, the listening CX and CN features explored in this work do not appear to have an effect on the DRL model for the prediction of music skips. Our analysis also reveals a temporal data leakage problem derived from some features in the dataset and used in the public challenge, since they provide information from the future that should not be made available to a live predictive system. Overall, this work shows that an accurate representation of the users' skipping behaviour can be achieved by leveraging a limited amount of user data. This offers strong implications for the design of novel user-centred MRSs with a minimisation and selection of high-quality data features to avoid introducing errors and biases. The results and a thorough analysis of our proposed approach indicate sample efficiency, swift convergence, and long-term stability of our proposed approach. With convergence reached using a significantly lower number of episodes, training time can be greatly reduced by early termination. With no GPU access required (in contrast to the state-of-the-art deep learning based models), our approach also clearly addresses the well-known limitation of DRL being a computationally extensive approach. These findings and the consistent performance with no signs of instability make this work of great interest for future research.

With the importance of modelling and understanding the users' skipping behaviour, we believe this work to be an important step towards improving user modelling techniques. An accurate representation of the skipping behaviour can provide an invaluable stream of information to the underlying recommendation process. For example, we expect our findings, e.g. the \textit{RS} type, to be highly relevant in the downstream task of capturing, in real-time, a user’s skipping type \cite{meggetto2021skipping}. By extending our approach to predict and understand other users' behaviours, we can create a holistic representation of the listeners' preferences, interests, and needs. We also advocate for thoughtful considerations when collecting and then presenting data to a model for measuring user behaviours. With increasingly rising concerns around users' data collection and privacy, the need for minimal data collection is paramount. Our proposed approach can be extended in future works to predict \textit{when} the song is likely to be skipped. This level of information could allow to predict moments in a song where skips are most likely to occur, which could be of great value for the underlying platform. Considering \textit{how} user's emotions or current psychological state affect their skipping behaviour is also an interesting venue for further research. With access to richer behavioural data and non-anonymised listening sessions, another line of research can investigate the relation between skipping signal and the individual user’s preferences (e.g., situation-aware MRS). Finally, although not the aim of this work, performance improvements are to be expected by further tailoring our approach to the music skip prediction task. Given the user-based exploratory nature of this work, we leave further experimentation and evaluations with emerging DRL model-free offline algorithms and architectures (e.g., extending our analysis to transformer-based DRL models \cite{janner2021offline}) for future investigation.

\begin{acks}
This work was supported by the Engineering and Physical Sciences Research Council [grant number EP/R513349/1].
\end{acks}

\bibliographystyle{ACM-Reference-Format}
\bibliography{paper}

\end{document}